\documentclass[pra,twocolumn,showpacs,preprintnumbers,amsmath,amssymb,superscriptaddress,unsortedaddress]{revtex4}
\pdfoutput=1
\usepackage{color}
\usepackage{graphicx}% Include figure files
\usepackage{dcolumn}% Align table columns on decimal point
\usepackage{bm}% bold math

\def\<{\langle}
\def\>{\rangle}
\begin{document}

\preprint{}

\title{Postulates for measures of genuine multipartite correlations}

\author{Charles H. Bennett}

\affiliation{IBM T. J. Watson Research Center, Yorktown Heights, New York 10598, USA}

\author{Andrzej Grudka}

\affiliation{Institute of Theoretical Physics and Astrophysics,
University of Gda\'{n}sk, 80-952 Gda\'{n}sk, Poland}

\affiliation{Faculty of Physics, Adam Mickiewicz University, 61-614
Pozna\'{n}, Poland}

\author{Micha{\l} Horodecki}

\affiliation{Institute of Theoretical Physics and Astrophysics,
University of Gda\'{n}sk, 80-952 Gda\'{n}sk, Poland}

\author{Pawe{\l} Horodecki}

\affiliation{Faculty of Applied Physics and Mathematics, Technical
University of Gda\'{n}sk, 80-952 Gda\'{n}sk, Poland}

\author{Ryszard Horodecki}

\affiliation{Institute of Theoretical Physics and Astrophysics,
University of Gda\'{n}sk, 80-952 Gda\'{n}sk, Poland}

\date{\today}% It is always \today, today,

\begin{abstract} A lot of research has been done on multipartite correlations. However, it seems strange that there is no definition of so called genuine multipartite correlations. In this paper we propose three reasonable postulates which each measure or indicator of genuine multipartite  correlations (or genuine multipartite  entanglement) should satisfy. We also introduce degree of correlations which gives partial characterization of multipartite correlations. Then, we show that covariance does not satisfy two postulates and hence, it cannot be used as an indicator of genuine multipartite correlations. Finally, we propose candidate for a measure of genuine multipartite correlations based on the work that can be drawn from local bath by means of a multipartite state.
\end{abstract}

\maketitle

\section{Introduction}

One of the most important problems in quantum information theory is the problem of quantifying correlations. Henderson and Vedral raised the problem of separating total correlations in a bipartite state into quantum and classical parts \cite{Henderson1} (see also in this context \cite{Ollivier, SaiToh, Modi}).They also proposed a measure of purely classical bipartite correlations . It was shown in \cite{Hayden1} that there exist bipartite states which have almost maximal entanglement of formation and almost no mutual information and hence, almost no classical correlations. In a series of papers a thermodynamical approach to quantifying correlations was developed \cite{Oppenheim1, Zurek, Horodecki3}. It is well known that bits of information can be used to extract work from a heat bath \cite{Bennett82}. If we have a bipartite quantum state one can ask how much work can be extracted from the heat bath under different classes of operations. In particular quantum information deficit was defined. It is the difference between globally and locally (with the use of local operations and classical communication) extractable work from the heat bath. Recently it was shown that under some restricted scenario of work extraction there exist quantum states for which quantum information deficit is equal to quantum mutual information \cite{Pankowski1}. As a result,  all correlations behave, as if they were exclusively quantum. 

The problem of coexistence of quantum and classical correlations in multipartite  systems  was considered in \cite{Kaszlikowski}. It was shown that there exist $n$-qubit states which consist of an equal mixture of two W states with an odd number of qubits, for which $n$-party covariance defined as $\text{Cov}(X_{1},... X_{n})=\langle (X_{1}-\langle X_{1}\rangle)... (X_{n}-\langle X_{n}\rangle) \rangle$ is zero for all choices of local observables $X_{i}$ and the state is genuinely entangled. Based on these observations the authors argued that genuine multipartite  correlations can exist without classical correlations. However, the conclusion is based on assumption that covariance is an indicator of genuine multipartite  correlations.

In this paper we formulate three postulates which any measure or indicator of genuine multipartite correlations should satisfy. We show that covariance does not satisfy two postulates when applied to more than two-partite systems. Our counterexamples will be the states considered in \cite{Kaszlikowski}.
Hence, vanishing of covariance for multipartite states does not imply absence of genuine multipartite classical correlations but rather shows that covariance cannot be an indicator of genuine multipartite classical correlations. As a by-product we obtain a protocol 
of distillation of W states from a wide class of states
(for distillation of W states from generic states see  \cite{Miyake}).

The paper is organized as follows. In Section II, we formulate reasonable postulates for measures or indicators of  genuine multipartite correlations and introduce degree of correlations.  In Section III we show that covariance does not satisfy postulates. In Section IV we discuss relation between multipartite correlations and work extraction. In Section V we draw our conclusions.

\section{Genuine $n$-partite correlations}

We do not know what it means that a state has genuine multipartite correlations. Hence we give reasonable postulates which each measure or indicator of genuine multipartite correlations should satisfy. In Subsection A we formulate postulates for genuine $n$-partite correlations of an $n$-partite state while in Subsection B we  formulate postulates for genuine $n$-partite correlations of an arbitrary multipartite state and introduce degree of correlations as an indicator of genuine $n$-partite correlations for an arbitrary multipartite state. 
The postulates  apply to correlations in general, hence, in particular, they apply also to genuine multipartite entanglement or genuine multipartite  classical correlations.

\subsection{Genuine $n$-partite correlations of an $n$-partite state}

Each measure or indicator of genuine $n$-partite correlations for an $n$-partite state should satisfy the following postulates:

{\bf Postulate 1.} \emph{If an $n$-partite state does not have genuine $n$-partite correlations and one adds a party in a product state then the resulting $n+1$ partite state does not have genuine $n$-partite correlations.}

{\bf Postulate 2.} \emph{If an $n$-partite state does not have genuine $n$-partite correlations then local operations and unanimous postselection
(which mathematically correspond to the operation $\Lambda_{1} \otimes \Lambda_{2} \otimes ... \otimes \Lambda_{n}$ where $m$ is the number of particles and 
each $\Lambda_{i}$ is trace non-increasing operation acting on $i$-th party's subsystem) cannot generate genuine $n$-partite correlations.}

{\bf Postulate 3.} \emph{If an $n$-partite state does not have genuine $n$-partite correlations then if one party splits his system into two, i.e. sends part of his system to new party which is product with the remainder of the system then the resulting $n+1$-partite state does not have genuine $n+1$-partite correlations.}

One can also require (compare Postulate 2) for each measure $C(\rho)$ of genuine $n$-partite correlations that it does not increase on average under local opertions, i.e.

\begin{eqnarray}
& C(\rho) \geq \sum_{i_1, i_2, ..., i_n}  p_{i_1, i_2, ..., i_n}\nonumber\\
& C(E^1_{i_1}  E^2_{i_2}... E^n_{i_n} \rho E^{1\dagger}_{i_1}  E^{2 \dagger}_{i_2}... E^{n \dagger}_{i_n}/p_{i_1, i_2, ..., i_n})
\end{eqnarray}
where $E^j_{i}$ are Krauss operators acting on $j$-th subsystem satisfying  $\sum _i E^{j \dagger}_{i}E^j_{i} \leq I$ and $p_{i_1, i_2, ..., i_n} =Tr(E^1_{i_1} \otimes E^2_{i_2} \otimes ... \otimes E^n_{i_n} \rho E^{1\dagger}_{i_1} \otimes E^{2 \dagger}_{i_2} \otimes ... \otimes E^{n \dagger}_{i_n})$. It seems that this requirement should be added to postulates of Henderson and Vedral for measures of classical bipartite correlations \cite{Henderson1}.

Let us now propose a definition of genuine multipartite of correlations.

{\bf Definition 1.}
\emph{A state of $n$ particles has genuine $n$-partite correlations if it is non-product in every bipartite cut.}

Below we prove that genuine multipartite correlations defined above satisfy Postulates 1-3.

{\bf Observation 1.} \emph{If genuine $n$-partite correlations are defined as in Definition 1 then they satisfy Postulates 1-3.}

\emph{Proof.} It is clear that genuine $n$-partite correlations satisfy Postulate 1. 

To show that they satisfy Postulate 2 we observe that an $n$-partite state which does not have genuine $n$-partite correlations is of the form
\begin{eqnarray}
\rho=\rho^{(n_1)}\otimes\rho^{(n_2)},
\end{eqnarray}
where $\rho^{(n_1)}$ and $\rho^{(n_2)}$  are states of $n_1$ and $n_2$ particles respectively ($n_1 + n_2=n$). It is clear that this product form is preserved by the operation $\Lambda_{1} \otimes \Lambda_{2} \otimes ... \otimes \Lambda_{n}$.

To show that genuine  $n$-partite correlations satisfy Postulate 3 we observe that a state which does not have genuine $n$-partite correlations after sending part of a one party's system is of the following forms

\begin{eqnarray}
\rho=\rho'^{(n_1+1)}\otimes\rho'^{(n_2)},
\end{eqnarray}
where $\rho'^{(n_1+1)}$, $\rho'^{(n_2)}$, are states of $n_1+1$ and $n_2$ particles respectively or

\begin{eqnarray}
\rho=\rho'^{(n_1)}\otimes\rho'^{(n_2+1)},
\end{eqnarray}
where $\rho'^{(n_1)}$, $\rho'^{(n_2+1)}$ are states of $n_1$, $n_2+1$ particles respectively. Hence, we see that it does not have genuine $n+1$-partite correlations.

{\bf Remark.} The postulates apply also to entanglement. However, the 
Definition 1 does not. More precisely, it does apply to pure state 
entanglement, i.e. one can say that pure state has genuine multipartite 
entanglement iff it is nonproduct with respect to any bipartite cut. To obtain 
definition of genuine multipartite entanglement for mixed states, we proceed in a standard way \cite{Werner1989}, i.e. we say that a state $\rho$ has 
genuine $n$-partite entanglement, if it is not a mixture of pure states that do not 
have genuine $n$-partite entanglement. Thus, to rule out 
correlations that do not represent entanglement, we can add fourth postulate,
saying, that by mixing states which do not have $n$-partite entanglement, 
we cannot obtain genuine $n$-partite entanglement.

\subsection{Degree of correlations}

We introduce the concept of degree of correlations. We do not define it first but rather require that it should satisfy the following postulates:

{\bf Postulate 1$'$.} \emph{If one adds a party in a product state with the remainder of the system then degree of correlations cannot change.}

{\bf Postulate 2$'$.} \emph{Local operations and postselection cannot increase degree of correlations. In particular local unitary operations cannot change degree of correlations.}

{\bf Postulate 3$'$.} \emph{If one party splits his system into two, i.e. sends part of his system to new party who is not correlated with the remainder of the system, then degree of correlations can increase at most by 1.}

Let us now propose a definition of degree of correlations.

{\bf Definition 2.} \emph{A state has degree of correlations equal to $n$ if there exists a subset of $n$ particles which has genuine $n$-particle correlations and there does not exist a subset of $m$ particles which has genuine $m$-particle correlations for any $m>n$.}

{\bf Example 1.}
An $n$-partite state of the form
\begin{equation}
\varrho=\frac{1}{2}\sum_{i=0}^{1}|ii...i\rangle \langle ii...i|
\end{equation}
has $2$-, $3$-, ...,$n$-partite correlations and degree of correlations equal to $n$.

{\bf Example 2.}
An $n$-partite state of the form
\begin{equation}
\varrho=\frac{2}{n}\sum_{i_{1}+i_{2}+...+i_{n}=0\mod 2}^{1}|i_{1}i_{2}...i_{n}\rangle \langle i_{1}i_{2}...i_{n}|
\end{equation}
has only $n$-partite correlations and degree of correlations equal to $n$. It does not have $2$-, $3$-, ...,$n-1$-partite correlations.

We can now find the form of a state which has degree of correlations equal to $n$.

{\bf Observation 2.} 
\emph{A state which has degree of correlations equal to $n$ is of the form:
\begin{eqnarray}
\label{form}
\rho=\rho^{(n)}\otimes\rho^{(m_1)}\otimes ... \otimes\rho^{(m_M)},
\end{eqnarray}
where $\rho^{(n)}$, $\rho^{(m_1)}$,..., $\rho^{(m_M)}$ are states of $n$, $m_1$,..., $m_M$ particles which are non-product in any bipartite cut and $n \geq m_{1}, ..., m_{M}$.}

{\it Proof.} It is clear that arbitrary state can be written in this form for some $n$. The state has degree of correlations at least equal to $n$ because the reduced state of first $n$ particles is non-product in any bipartite cut. To show that degree of correlations cannot be greater than $n$ it is enough to notice that if we trace some particles then if the state is product in some cut before we trace particles then it will be product in this cut after we trace particles.

Below we prove that the degree of correlations defined above satisfies Postulates 1$'$-3$'$.

{\bf Observation 3.} \emph{If  genuine $n$-partite correlations and degree of correlations are defined as in Definition 1 and Definition 2 then degree of correlations satisfies Postulates 1$'$-3$\,'$.}

\emph{Proof.} It is clear that degree of correlations satisfies Postulate 1$'$. 

To show that it satisfies Postulate 2$'$  we use the fact proved in Observation 1 that a state which has degree of correlations equal to $n$ is of the form:
\begin{eqnarray}
\label{form}
\rho=\rho^{(n)}\otimes\rho^{(m_1)}\otimes ... \otimes\rho^{(m_M)},
\end{eqnarray}
where $\rho^{(n)}$, $\rho^{(m_1)}$,..., $\rho^{(m_M)}$ are states of $n$, $m_1$,..., $m_M$ particles respectively which are non-product in any bipartite cut and $n \geq m_{1}, ..., m_{M}$. It is clear that this product form is preserved by the operation $\Lambda_{1} \otimes \Lambda_{2} \otimes ... \otimes \Lambda_{k}$, where $k=n+m_1+...+m_M$ is the number of parties.

To show that degree of correlations satisfies Postulate 3$'$ we observe that a state which has degree of correlations equal to $n$ after sending part of a one party's system is of the following form

\begin{eqnarray}
\rho=\rho'^{(n+1)}\otimes\rho'^{(m_1)}\otimes ... \otimes\rho'^{(m_M)},
\end{eqnarray}
where $\rho'^{(n+1)}$, $\rho'^{(m_1)}$,..., $\rho'^{(m_M)}$ are states of $n+1$, $m_1$,..., $m_M$ particles or

\begin{eqnarray}
\rho=\rho'^{(n)}\otimes\rho'^{(m_1+1)}\otimes ... \otimes\rho'^{(m_M)},
\end{eqnarray}
where $\rho'^{(n)}$, $\rho'^{(m_1+1)}$,..., $\rho'^{(m_M)}$ are states of $n$, $m_1+1$,..., $m_M$ particles and so on. Hence, we see that it has degree of correlations at most $n+1$.

Using Postulates 1-3 we can prove the following Observation.

{\bf Observation 4.} \emph{If the initial state has degree of correlations less than $n$ then if the parties add $k$ ancillas in a product state, perform local operations on their particles and their ancillas, and send ancillas to $k$ new parties in such a way that each new party receives an ancilla from only one party then the final state cannot have degree of correlations greater or equal to $n+k$.}

{\it Proof.} First, $k$ parties add ancillas in a product state. Degree of correlations cannot change (Postulate 1$'$). Next, $k$ parties apply local operations to their original qubits and added ancillas. Degree of correlations cannot increase (Postulate 2$'$). Finally, $k$ parties send ancillas to $k$ new parties. Degree of correlations can increase at most by $k$ (it can increase by $1$ for each sent ancilla) (Postulate 3$'$).

\begin{figure}
\label{fig:1a}
\hspace{-4cm} a) \\
\hspace{-8mm}\includegraphics [width=2truecm] {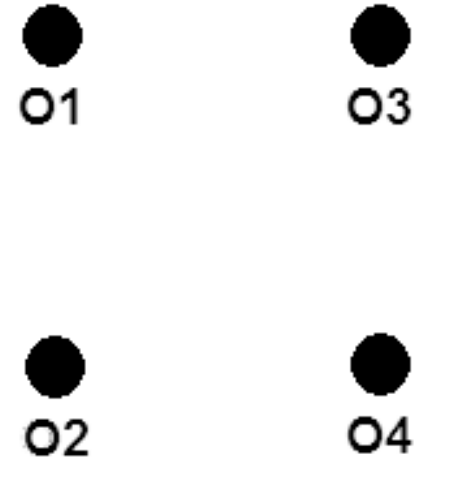}\\
\hspace{-4cm} b) \\
\includegraphics [width=3truecm] {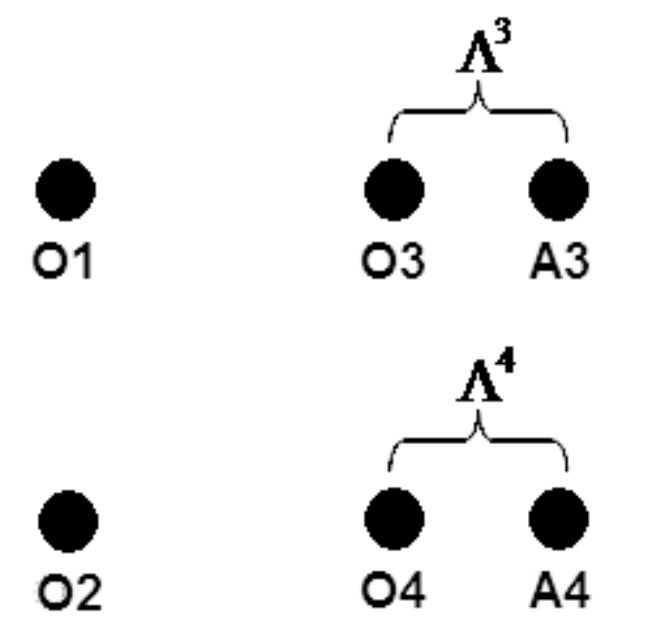}\\
\hspace{-4cm} c) \\
\hspace{7mm}\includegraphics [width=3.5truecm] {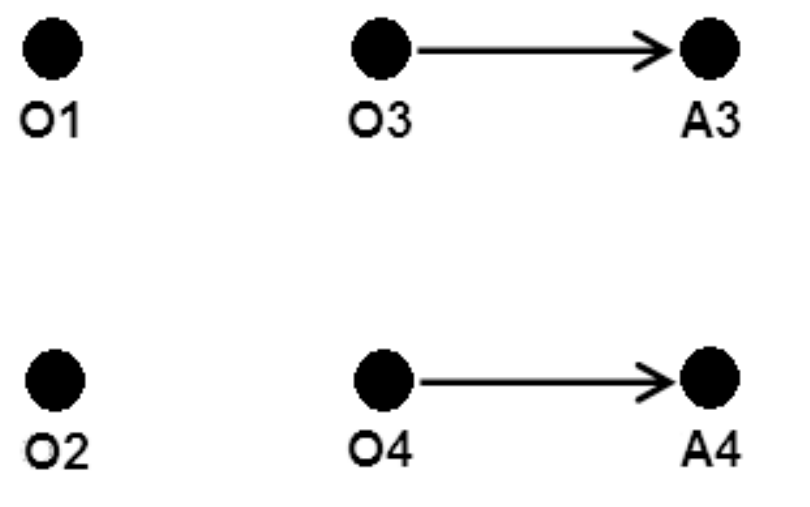}\\
\caption{Illustration of Observation 1. (a) Four parties share a joint state $\rho$. (b1) The third and the forth party add ancillas in product state which is product with the original system and the other ancilla. After this step the parties share the state $\rho \otimes \rho_{A3} \otimes \rho_{A4}$. (b2) The third and the fourth party perform local operations on their original qubits and ancillas. After this step the parties share the state $\text{id}_{O1} \otimes \text{id}_{O2} \otimes \Lambda_{O3A3}^1  \otimes \Lambda_{O4A4}^2(\rho \otimes \rho_{A3} \otimes \rho_{A4})$, where $\text{id}_{i}$ denotes the identity map acting on qubit $i$ and $\Lambda_{ij}^a$  denotes any completely positive map acting on qubits $i$ and $j$. (c) The third and the fourth party send ancillas to the fifth and sixth party, respectively. If the initial state has degree of correlations less than $n$ then the final state cannot have degree of correlations greater or equal to $n+2$.
}
\end{figure}

\section{Covariance does not satisfy postulates}

\subsection{Postulate 2}

Before we show that covariance does not satisfy Postulate 2 we present a purification protocol which allows the distillation of W states from certain mixed state. The protocol consists of two steps. In the first step each party performs a measurement on his particle -- the so called local filtering \cite{Gisin, Bennett2, Horodecki2}. The measurements performed by all parties are independent, i.e. they are not conditioned on the results of the measurements performed by other parties. In the second step the parties postselect a state. 
The postselected state can have fidelity with W state as close to $1$ as one wants. However, the probability of distilling such a state decreases with fidelity. This protocol is a multipartite  version of the so called quasi-distillation process \cite{ Horodecki4, Brassard}. 

Let us consider a state which is a mixture of W state and a normalized state $\rho^{(1)}$ with support contained in the $2^n-n-1$-dimensional Hilbert space spanned by all vectors which have $2$ or more $1$'s, i.e.
\begin{eqnarray}
\rho=p|W\rangle \langle W|+(1-p)\rho^{(1)}.
\end{eqnarray}
Let each party perform a measurement described by the following Kraus operators
\begin{eqnarray}
E_{S}=|0\rangle \langle0|+\sqrt{\epsilon} |1\rangle \langle1| \nonumber \\
E_{F}=\sqrt{1-\epsilon}|1\rangle \langle 1|
\end{eqnarray}
The action of $E_{S}^{\otimes n}$ on states with $m$ $1$'s and $n-m$ $0$'s is given by the following formula
\begin{eqnarray}
E_{S}^{\otimes n}|1\rangle^{\otimes m}|0\rangle^{\otimes n-m}=\sqrt{\epsilon}^m|1\rangle^{\otimes m}|0\rangle^{\otimes n-m}
\end{eqnarray}
A similar result applies for all permutations of $m$ $1$'s and $n-m$ $0$'s.

Hence, if each party obtains $S$ as the result of the measurement then the post-measurement state is proportional to
\begin{eqnarray}
\rho'=E_{S}^{\otimes n} \rho E_{S}^{ \otimes n}=\epsilon p|W \rangle \langle W|+\epsilon^2(1-p)\rho'^{(1)},
\end{eqnarray}
where $\rho'^{(1)}$ is unnormalized state with the sum of eigenvalues less or equal to $1$ and with support contained in the $(2^n-n-1)$-dimensional Hilbert space orthogonal to the W state.
The probability that each party obtains $S$ as the result of the measurement is 
\begin{eqnarray}
q=\text{Tr}(\rho')>\epsilon p. 
\end{eqnarray}
The fidelity of the post-measurement state with the W state is:
\begin{eqnarray}
F=\frac{\langle W|\rho'|W \rangle}{\text{Tr}(\rho')}>\frac{\epsilon p}{\epsilon p+\epsilon^2(1-p)}
\end{eqnarray}
If $\epsilon$ is small then the fidelity is close to $1$. Hence, with small probability it is possible to distill a state close to $W$.
More precisely, the probability that we distill state W with fidelity $F$ satisfies
\begin{eqnarray}
q > \frac{p^2}{(1-p)}\frac{1-F}{F}
\end{eqnarray}

Consider now as example the $n$-partite state from \cite{Kaszlikowski} which is an equal mixture of $W$ and $\overline{W}$ states consisting of an odd number of qubits, i.e.,
\begin{eqnarray}
\label{mixture}
\rho=\frac{1}{2}|W\rangle \langle W|+\frac{1}{2}|\overline{W}\rangle \langle \overline{W}|,
\end{eqnarray}
where
\begin{eqnarray}
|W\rangle=\frac{1}{\sqrt{n}}(|10...0\rangle+|01...0\rangle+...+|00...1\rangle)
\end{eqnarray}
and
\begin{eqnarray}
|\overline W\rangle=\frac{1}{\sqrt{n}}(|01...1\rangle+|10...1\rangle+...+|11...0\rangle)
\end{eqnarray}
One can show that for this state all $n$-partite covariances vanish \cite{Kaszlikowski} (For completeness we show it in Appendix A.).

Let the parties apply just described protocol to the state of Eq. (\ref{mixture}) mixtures of two W states. If each party obtains $S$ as the result of the measurement then the state is proportional to
\begin{eqnarray}
\rho'=\epsilon \frac{1}{2}|W\rangle \langle W|+\epsilon^{n-1}\frac{1}{2}|\overline{W}\rangle \langle \overline{W}|
\end{eqnarray}
The probability that each party obtains $S$ as the result of measurement is 
\begin{eqnarray}
q=\frac{1}{2}\epsilon(1+\epsilon^{n-2})
\end{eqnarray}
The fidelity of the post-measurement state with W state is:
\begin{eqnarray}
F=\frac{\langle W|\rho'|W \rangle}{\text{Tr}(\rho')}=\frac{1}{1+\epsilon^{n-2}}
\end{eqnarray}
Let us take $\epsilon=1-\frac{1}{\sqrt{n}}$ and calculate the limit of $q$ and $F$ for large $n$. We have
\begin{eqnarray}
q \approx \frac{1}{2}(1-\frac{1}{\sqrt{n}})(1+e^{-\sqrt{n}})
\end{eqnarray}
and
\begin{eqnarray}
F \approx \frac{1}{1+e^{-\sqrt{n}}}
\end{eqnarray}
More generally, the probability that we distill the state W with fidelity $F$ is:
\begin{eqnarray}
q=\frac{1}{2F} \left( \frac{1-F}{F} \right)^\frac{1}{n-2}
\end{eqnarray}

In Fig. 2, we show how the probability of distillation depends on the fidelity of the distilled state for mixtures of two W states. For large $n$, the probability of obtaining a state close to W is close to $\frac{1}{2}$. Hence, asymptotically the measurement \emph{effectively projects} the initial state on W state.

\begin{figure}
\includegraphics {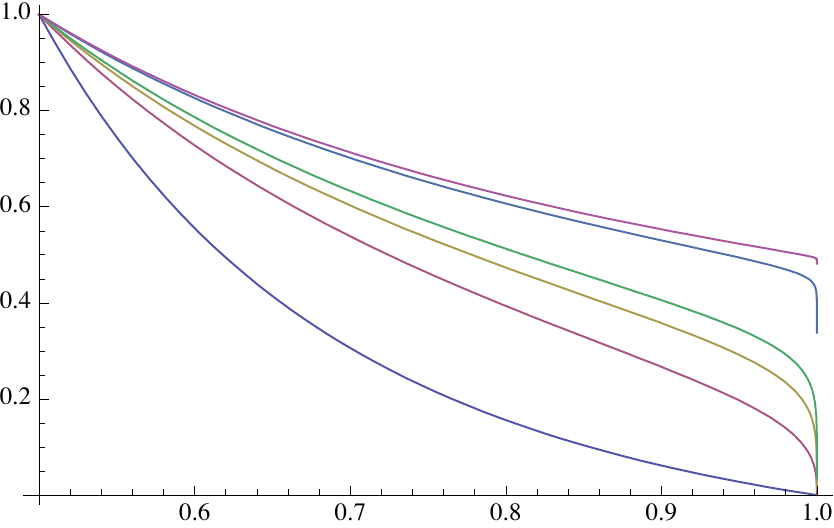}
\caption{\label{fig:2} Probability of distillation versus fidelity. From bottom to top: 3-qubit state, 5-qubit state, 7-qubit state, 9-qubit state, 49-qubit state and 499-qubit state.}
\end{figure}

We have thus shown that we can transform with local operations and postselection the state of Eq. (\ref{mixture}) into a state of the form
\begin{eqnarray}
\rho=F|W\rangle \langle W|+(1-F)|\overline{W}\rangle \langle \overline{W}|,
\end{eqnarray}
where $0 < F < 1$ can be arbitrary close to 1. The $n$-partite covariance $\text{Cov}(\sigma_z^{1},... \sigma_z^{n})$ of the above 
state is given by the following expression (we show it in Appendix A).

\begin{eqnarray}
& \text{Cov}(\sigma_z^{1},... \sigma_z^{n})=\text{Tr}((\sigma_z-\langle\sigma_z\rangle)^{\otimes n}\rho)= \nonumber \\
& =-F (1-\langle\sigma_z\rangle)^{n-1}(1+\langle\sigma_z\rangle) + \nonumber \\
& +(1-F)(-1)^{n-1}(1+\langle\sigma_z\rangle)^{n-1}(1-\langle\sigma_z\rangle),
\end{eqnarray}
where
\begin{eqnarray}
&\langle\sigma_z\rangle=\text{Tr}(\sigma_z \rho)= \nonumber \\
& = F\text{Tr}(\sigma_z |W \rangle\langle W|)+(1-F)\text{Tr}(\sigma_z |\overline{W}\rangle\langle \overline{W}|)= \nonumber \\
& =(2F-1)\frac{n-2}{n}.
\end{eqnarray}

In Fig. 3, we present how it depends on the fidelity $F$ for a 3-qubit state and a 9-qubit state. For both states the covariance vanishes only for $F=\frac{1}{2}$.

Let us summarize this result. The parties start with a state for which all covariances vanish. Then they apply local filtering and postselect a state. They can choose measurements in such a way that the postselected state has nonvanishing covariance. Hence, we have shown that covariance does not satisfy our second postulate. This once again shows that covariance should not be regarded as an indicator of genuine multipartite classical correlations.

\begin{figure}
\includegraphics {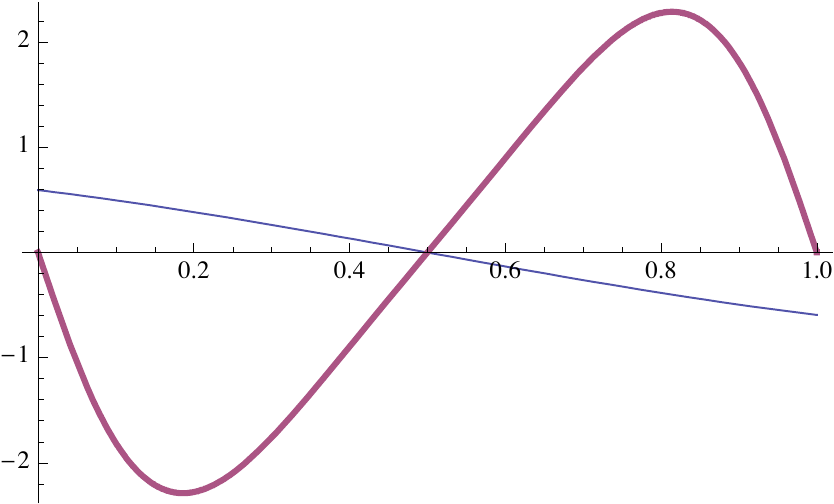}
\caption{\label{fig:3} Covariance versus fidelity for 3-qubit state (thin line) and 9 qubit state (thick line)}
\end{figure}

\subsection{Postulate 3}

Let $n$ parties share a state of Eq. (\ref{mixture}). Let each party add to his original qubit an auxiliary qubit in state $| 0 \rangle$, performs CNOT gate (the original qubit is control qubit and the auxiliary qubit is target qubit) and send the ancilla to new party. Each new party receives an ancilla from only one party. The $2n$-partite state is

\begin{eqnarray}
\rho'=\frac{1}{2}|W'\rangle \langle W'|+\frac{1}{2}|\overline{W}'\rangle \langle \overline{W}'|,
\end{eqnarray}
where
\begin{eqnarray}
& |W'\rangle=\frac{1}{\sqrt{n}}(|10...0;10...0\rangle+ \nonumber \\
& +|01...0;01...0\rangle+...+|00...1;00...1\rangle)
\end{eqnarray}
and
\begin{eqnarray}
& |\overline{W}'\rangle=\frac{1}{\sqrt{n}}(|01...1;01...1\rangle+ \nonumber \\
& +|10...1;10...1\rangle+...+|11...0;11...0\rangle)
\end{eqnarray}
One can show that the following covariance
\begin{eqnarray}
C=\text{Tr}((\sigma_z-\langle\sigma_z\rangle)^{\otimes 2n}\rho'),
\end{eqnarray}
is equal to 1. We see that covariance does not satisfy Postulate 3 when applied to more than two-partite states. 

\section{Multipartite correlations and work extraction}

In this section we investigate the relation between multipartite correlations and the amount of work that can be extracted from the environment \cite{Bennett82,Oppenheim1, Horodecki3,silnik}. Let us suppose that two parties share a quantum state $\rho_{AB}$. It is well known that this state can be used to extract work from the environment in many different ways. We consider two scenarios. In the first one the parties are allowed to perform closed local operations, i.e., they can perform local unitary operations and local dephasing (CLO). In the second one they are allowed to perform closed local operations and send classical communication, i.e. send subsystem down a completely dephasing channel (CLOCC). If for the state $\rho_{AB}$ the parties can extract more work with CLOCC than with CLO then the state $\rho_{AB}$ has classical correlations. Let us now consider multipartite states. By analogy we expect that if the parties can extract more work with CLOCC and with sending classical information across any bipartite cut than with CLOCC and without sending classical information across at least one cut then the state has genuine multipartite classical correlations. The  difference between extractable work in those two scenarios minimized overall bipartite cuts we shall denote by $\delta W$.

{\bf Example 1.}
Here we compare the amount of work extractable with the use of an equal mixture of two tripartite $W$ states when three parties cooperate, i.e., there is classical communication across any cut, and when two parties cooperate, i.e., there is not classical communication across one cut. Let us first suppose that three parties cooperate. If the parties dephase their qubits in $\{ |0\rangle, |1\rangle \}$ basis and the first two parties send their qubits to the third party (This is equivalent to sending qubits down a completely dephasing channel) then the third party will hold the state
\begin{eqnarray}
& \sigma_{123}=\frac{1}{6}(|001\rangle \langle 001|+|010\rangle \langle 010|+|100\rangle \langle 100|+\nonumber\\
& +|110\rangle \langle 110|+|101\rangle \langle 101|+|011\rangle \langle 011|).
\end{eqnarray}
He can now extract $3-\log_2 6\approx 0.4150$ bits of work.
Another protocol is the following: one party measures in a chosen basis,
and tells the result to other parties, who then draw work from the resulting pure state they share. If the basis is $|0\>,|1\>$, the result is the same as above. The complementary basis $|+\>,|-\>$ gives $0.4499$, while the optimal basis is $\sqrt{1\over 3}|0\>+\sqrt{2\over 3}|1\>, \sqrt{2\over 3}|0\>-\sqrt{1\over 3}|0\>$, providing $0.4502$ bits of work. We do not know, whether by general CLOCC protocol one can extract more work.

Let us now suppose that only two parties cooperate, i.e. the first and the second one. The reduced state of the third party is maximally mixed state and he cannot extract any work at all. The reduced state of the first two parties is Bell diagonal state
\begin{eqnarray}
& \rho_{12}=\text{Tr}_3\rho=\frac{2}{3}|\Psi^+\rangle \langle \Psi^+|\nonumber\\
&+\frac{1}{6}|\Phi^+\rangle \langle \Phi^+|+\frac{1}{6}|\Phi^-\rangle \langle \Phi^-|
\end{eqnarray}
where
\begin{eqnarray}
& |\Psi^+\rangle=\frac{1}{2}(|01\rangle+|01\rangle)\nonumber\\
& |\Phi^\pm\rangle=\frac{1}{2}(|00\rangle \pm |11\rangle)
\end{eqnarray}
If the first party dephase his qubit in $\{ |+\rangle, |-\rangle \}$ basis, where
\begin{eqnarray}
|\pm\rangle=\frac{1}{\sqrt{2}}(|0\rangle \pm |1\rangle)
\end{eqnarray}
and send his qubit to the second party then the second party after applying local unitary operation will hold the state
\begin{eqnarray}
& \sigma_{12}=\frac{5}{12} |01\rangle \langle 01|+\frac{5}{12}|10\rangle \langle 10|+\nonumber\\
&+\frac{1}{12}|00\rangle \langle 00|+\frac{1}{12}|11\rangle \langle 11|
\end{eqnarray}
He can now extract $1-H(\frac{5}{6})\simeq 0.3499$ bits of work, where $H(x)=-x\log_2 x -(1-x)\log_2 (1-x)$ is binary entropy. This is maximal work which can be extracted with help of one-way classical communication even if one takes into account 
POVMS and asymptotic limit of many copies \cite{Horodecki3}. However, again, it is not known if one can extract more work with help of two-way classical communication.

To summarize, we showed, that if the communication through A:BC cut 
is allowed, then we can extract at least $0.4502$ bits of work, while if it is not allowed, we were able to provide a protocol which extracts $0.3499$ bits of work
(this concerns all possible cuts, as the state is permutationally symmetric). 
If the latter protocol were optimal, we would have $\delta W \gtrsim 0.1$.
This supports existence of genuine tripartite correlations in the state. 

{\bf Example 2.}
On the other hand if the parties can extract the same amount of work with CLOCC and with sending classical information across any bipartite cut than with CLOCC and without sending classical information across at least one cut then we cannot conclude that the state does not have genuine multipartite classical correlations. Let us consider the following tripartite state
\begin{eqnarray}
\rho_{123}=|\Psi^+\rangle \langle \Psi^+|_{12}|0\rangle \langle 0|_{3}+|\Psi^-\rangle \langle \Psi^-|_{12}|1\rangle \langle 1|_{3}
\end{eqnarray}
If three parties cooperate they can extract one bit of work. However if only the first and the second party cooperate they can also extract one bit of work. On the other hand the state is non-product across any bipartite cut and according to Definition 2 it has genuine tripartite classical correlations.

The above two examples show that on one hand $\delta W$ can indicate 
multipartite correlations, on the other hand it may vanish, even though 
the state is non-product against any cut. This is analogous to behavior 
of some entanglement measures: e.g. distillable entanglement can vanish 
for states despite they are entangled. Thus $\delta W$ quantifies some 
particular type of genuine multipartite correlations, which may be absent in 
some states even though they contain genuinely multipartite correlations.

Let us note that the above property of $\delta W$ is similar to covariance, which disappears for state \eqref{mixture}, even though it has  genuine multipartite correlations with respect to some other criteria. 
One basic difference is however, that covariance can be positive 
even for states such as product of EPR pairs $\Psi^+_{AB}\otimes \Psi^+_{CD}$ 
which quite obviously do not represent multipartite entanglement 
(example of such covariances are those of Pauli matrices in cases when pairs of the same 
Pauli matrices are measured on each of the EPR pair). Moreover, we believe  (though have not proven) that $\delta W$ would satisfy our postulates, i.e. having been zero for some state,
it will not go up under the operations described in the formulation of the postulates, while covariance, as we have shown in previous sections violates 
two postulates.

\section{Conclusions}

In conclusion we have proposed reasonable postulates which each measure or indicator of genuine multipartite  correlations (or genuine multipartite  entanglement) should satisfy. We also introduced degree of correlations which gives partial characterization of multipartite correlations. We have shown that covariance does not satisfy proposed postulates and it cannot be used as an indicator of genuine multipartite classical correlations. In particular, our postulates show that the claim that there exist genuine $n$-partite quantum correlations without genuine $n$-partite classical correlations is not justified. As a by-product we obtain a protocol  of distillation of W states from a wide class of states. Finally we propose a candidate for a measure of genuine multipartite correlations based on work that can be drawn  from local environments by means a multipartite state. 
We hope that our results, especially the proposed postulates, will allow to develop understanding and quantitative description of genuine multipartite  correlations. 

\begin{acknowledgments}
We thank R. W. Chhajlany, O. G\"uhne, J. Eisert, J. Mod{\l}awska, and M. Piani for discussions, and the authors of \cite{Kaszlikowski} for useful feedback.
This work was supported by the European Commission through the Integrated Project FET/QIPC ``SCALA'' and by Ministry of Science and Higher
Education grant N N202 231937. Part of this work was done in National Quantum Information Centre of Gda\'nsk.
\end{acknowledgments}

\appendix
\section{Calculation of covariance}

We calculate all covariances for an equal mixture of two W states consisting of an odd number of qubits
\begin{eqnarray}
\rho=\frac{1}{2}|W\rangle \langle W|+\frac{1}{2}|\overline{W}\rangle \langle \overline{W}|,
\end{eqnarray}
We can write it as
\begin{eqnarray}
& \text{Cov}(X_{1},... X_{n})=\text{Tr}((X_{1}-\langle X_{1}\rangle)...(X_{n}-\langle X_{n}\rangle)\rho)= \nonumber \\
& =F\langle W|(X_{1}-\langle X_{1}\rangle)...(X_{n}-\langle X_{n}\rangle)|W \rangle+ \nonumber \\
& +(1-F)\langle \overline{W}|(X_{1}-\langle X_{1}\rangle)...(X_{n}-\langle X_{n}\rangle)|\overline{W}\rangle,
\end{eqnarray}
where $X_{i}$ denotes Pauli matrix acting on $i$-th qubit. 
Since $\langle W|X_{i}|W\rangle=0$ we have
\begin{eqnarray}
& \text{Cov}(X_{1},... X_{n})=\nonumber \\
& \frac{1}{2}\langle W|X_{1}... X_{n}|W\rangle+\frac{1}{2}\langle \overline{W}|X_{1}... X_{n}|\overline{W}\rangle
\end{eqnarray}
and we need to calculate $\langle W|X_{1}... X_{n}|W\rangle$ and  $\langle \overline{W}|X_{1}... X_{n}|\overline{W}\rangle$. Moreover, since $\sigma_x$ and $\sigma_y$ exchange $|0\rangle$ and $|1\rangle$ the only non-vanishing terms are those which contain\\
1) $n$ times $\sigma_z$\\
2) $2$ times $\sigma_x$ and $n-2$ times $\sigma_z$\\
3) $2$ times $\sigma_y$ and $n-2$ times $\sigma_z$\\
4) $1$ time $\sigma_x$, $1$ time $\sigma_y$, and $n-2$ times $\sigma_z$\\
The other products of Pauli matrices when acting on $W$ or $\overline{W}$ state transform it into some orthogonal state, and hence their expectation value in $W$ or $\overline{W}$ state is equal to zero. We do not consider in detail the above four cases as their are similar and restrict our attention only to the second case. We have
\begin{eqnarray}
\sigma_x^1\sigma_x^2\sigma_z^{3}...\sigma_z^{n}|100...0\rangle=|010...0\rangle
\end{eqnarray}

\begin{eqnarray}
\sigma_x^1\sigma_x^2\sigma_z^{3}...\sigma_z^{n}|010...0\rangle=|100...0\rangle
\end{eqnarray}

\begin{eqnarray}
\sigma_x^1\sigma_x^2\sigma_z^{3}...\sigma_z^{n}|011...1\rangle=(-1)^n|101...1\rangle
\end{eqnarray}

\begin{eqnarray}
\sigma_x^1\sigma_x^2\sigma_z^{3}...\sigma_z^{n}|101...1\rangle=(-1)^n|011...1\rangle
\end{eqnarray}
and hence
\begin{eqnarray}
\langle W|\sigma_x^1\sigma_x^2\sigma_z^{3}...\sigma_z^{n}|W \rangle=\frac{2}{n}
\end{eqnarray}
\begin{eqnarray}
\langle  \overline{W}|\sigma_x^1\sigma_x^2\sigma_z^{3}...\sigma_z^{n}|  \overline{W} \rangle=(-1)^n\frac{2}{n}
\end{eqnarray}
We obtain
\begin{eqnarray}
\text{Tr}(\sigma_x^1\sigma_x^2\sigma_z^{3}...\sigma_z^{n}\rho)=0.
\end{eqnarray}
and similarly for other products of $n$ Pauli matrices.

We calculate the covariance $\text{Cov}(\sigma_z^{1},... \sigma_z^{n})$ for an arbitrary mixture of two W states consisting of an odd number of qubits
\begin{eqnarray}
\rho=F|W\rangle \langle W|+(1-F)|\overline{W}\rangle \langle \overline{W}|
\end{eqnarray}
We can write it as
\begin{eqnarray}
& \text{Cov}(\sigma_z^{1},... \sigma_z^{n})=\text{Tr}((\sigma_z-\langle\sigma_z\rangle)^{\otimes n}\rho)= \nonumber \\
& =F\text{Tr}((\sigma_z-\langle\sigma_z\rangle)^{\otimes n}|W \rangle\langle W|)+ \nonumber \\
& +(1-F)\text{Tr}((\sigma_z-\langle\sigma_z\rangle)^{\otimes n}|\overline{W}\rangle\langle \overline{W}|).
\end{eqnarray}
If we expand $(\sigma_z-\langle\sigma_z\rangle)^{\otimes n}$ in a series we obtain a term of the form $\sigma_z^1\sigma_z^{i_1}...\sigma_z^{i_{k-1}}\langle\sigma_z\rangle^{n-k}$ (i.e. one Pauli matrix acts on the first qubit and $k-1$ Pauli matrices act on some of the remaining qubits)$ \frac{(n-1)!}{(n-k)!(k-1)!}$ times. On the other hand we obtain a term of the form $\sigma_z^{i_1}...\sigma_z^{i_k}\langle\sigma_z\rangle^{n-k}$ (i.e. $k$ Pauli matrices act on some of the remaining qubits) $\frac{(n-1)!}{(n-1-k)!(k)!}$ times.

Since the state has high symmetry we can consider how the above operators act only on the states $|10...0\rangle$ and $|01...1\rangle$. We have
\begin{eqnarray}
\sigma_z^1\sigma_z^{i_1}...\sigma_z^{i_{k-1}}|10...0\rangle=-|10...0\rangle,
\end{eqnarray}
\begin{eqnarray}
\sigma_z^{i_1}...\sigma_z^{i_k}|10...0\rangle=|10...0\rangle,
\end{eqnarray}
\begin{eqnarray}
\sigma_z^1\sigma_z^{i_1}...\sigma_z^{i_{k-1}}|01...1\rangle=(-1)^{k-1}|01...1\rangle
\end{eqnarray}
and
\begin{eqnarray}
\sigma_z^{i_1}...\sigma_z^{i_k}|01...1\rangle=(-1)^k|01...1\rangle
\end{eqnarray}
where $i_1, ...,i_{k} \in \{2,...,n\}$.
We obtain
\begin{eqnarray}
& \text{Tr}((\sigma_z-\langle\sigma_z\rangle)^{\otimes n}|W \rangle\langle W|)= \nonumber \\
& =\sum_{k=0}^{n-1}\frac{(n-1)!}{(n-1-k)!k!}(-\langle\sigma_z\rangle)^{(n-k)}-\nonumber\\ 
& -\sum_{k=1}^{n}\frac{(n-1)!}{(n-k)!(k-1)!}(-\langle\sigma_z\rangle)^{(n-k)}
\end{eqnarray}
and
\begin{eqnarray}
& \text{Tr}((\sigma_z-\langle\sigma_z\rangle)^{\otimes n}|\overline{W}\rangle\langle \overline{W}|)= \nonumber \\
& =\sum_{k=0}^{n-1}\frac{(n-1)!}{(n-1-k)!k!}(-1)^k(-\langle\sigma_z\rangle)^{(n-k)}-\nonumber\\ 
& -\sum_{k=1}^{n}\frac{(n-1)!}{(n-k)!(k-1)!}(-1)^k(-\langle\sigma_z\rangle)^{(n-k)}.
\end{eqnarray}
After some algebra, we obtain
\begin{eqnarray}
& \text{Tr}((\sigma_z-\langle\sigma_z\rangle)^{\otimes n}|W \rangle\langle W|)= \nonumber \\
& =-(1-\langle\sigma_z\rangle)^{n-1}(1+\langle\sigma_z\rangle)
\end{eqnarray}
and
\begin{eqnarray}
& \text{Tr}((\sigma_z-\langle\sigma_z\rangle)^{\otimes n}|\overline{W}\rangle\langle \overline{W}|)= \nonumber \\
& =(-1)^{n-1}(1+\langle\sigma_z\rangle)^{n-1}(1-\langle\sigma_z\rangle).
\end{eqnarray}
We arrive at the following expression for covariance
\begin{eqnarray}
& \text{Cov}(\sigma_z^{1},... \sigma_z^{n})=\text{Tr}((\sigma_z-\langle\sigma_z\rangle)^{\otimes n}\rho)= \nonumber \\
& =-F (1-\langle\sigma_z\rangle)^{n-1}(1+\langle\sigma_z\rangle) + \nonumber \\
& +(1-F)(-1)^{n-1}(1+\langle\sigma_z\rangle)^{n-1}(1-\langle\sigma_z\rangle).
\end{eqnarray}
The average value of $\sigma_z$ for $W$ and $\overline{W}$ states is $\frac{n-2}{n}$ and $-\frac{n-2}{n}$, respectively and the average value of $\sigma_z$ for a mixture of $W$ and $\overline{W}$ states is
\begin{eqnarray}
&\langle\sigma_z\rangle=\text{Tr}(\sigma_z \rho)= \nonumber \\
& = F\text{Tr}(\sigma_z |W \rangle\langle W|)+(1-F)\text{Tr}(\sigma_z |\overline{W}\rangle\langle \overline{W}|)= \nonumber \\
& =(2F-1)\frac{n-2}{n}
\end{eqnarray}

\end{document}